\documentclass[twocolumn,prb,showpacs]{revtex4-1}

\usepackage[utf8]{inputenc}
\usepackage{mathtools}
\usepackage{amsmath,amssymb}    
\usepackage{graphicx}   
\usepackage{color}     
\DeclarePairedDelimiter\bra{\langle}{\rvert}
\DeclarePairedDelimiter\ket{\lvert}{\rangle}
\DeclarePairedDelimiterX\braket[2]{\langle}{\rangle}{#1 \delimsize\vert #2}
\DeclarePairedDelimiter\mean{\langle}{\rangle}
\newcommand*{\diff}{d}
\newcommand*{\tr}[1]{\mathrm{tr}\left(#1\right)}

\begin{document}

\title{Doublon lifetimes in dissipative environments}

\author{Miguel Bello}
\author{Gloria Platero}
\author{Sigmund Kohler}
\affiliation{Instituto de Ciencias de Materiales de Madrid, CSIC, E-28049, Spain}
\date{\today}

\begin{abstract}
We study the dissipative decay of states with a doubly occupied site in a
two-electron Hubbard model, known as doublons.  For the environment we
consider charge and current noise which are modelled as a bosonic
heat bath that couples to the onsite energies and the tunnel couplings,
respectively.  It turns out that the dissipative decay depends
qualitatively on the type of environment as for charge noise, the
life time grows with the electron-electron interaction.  For current noise,
by contrast, doublons become increasingly unstable with larger interaction.
Numerical studies within a Bloch-Redfield approach are complemented by
analytical estimates for the decay rates.  For typical quantum dot
parameters, we predict that the doublon life times up to 50\,ns.
\end{abstract}

\maketitle

\section{Introduction}

In recent years experiments with strongly-interacting cold atomic gases
have attracted much attention.\cite{ReviewBloch2008}  A particular
advantage of these systems is that their parameters can be controlled to a
high degree either directly or via oscillating forces that lead to
synthetic gauge fields.\cite{Creffield2011a, Dalibard2011a}  This allows a flexible
engineering and simulation of many-body Hamiltonians.  For a theoretical
description, one frequently employs the Hubbard model.  Despite its seeming
simplicity, it captures a great variety of condensed-matter phenomena
ranging from metallic behavior to insulators, magnetism, and
superconductivity.

In the strongly interacting limit of the Hubbard model, particles occupying
the same lattice site can bind together, even for repulsive interactions.
This occurs when the onsite interaction is much larger than the tunneling 
such that energy conservation inhibits the decay into a
state with two distant particles.  In principle, both bosons
\cite{Valiente2008, Compagno2017} and fermions \cite{BookEssler2005} can
form such $N$-particle states.  While the former allow any occupation
number, for fermions with spin $s$, the occupation of one site is
restricted to at most $2s+1$ particles. 
In particular, two spin-1/2 fermions may reside in a singlet spin 
configuration on one lattice site and, thus, form a doublon. Over the last 
years, they have been investigated both theoretically 
\cite{Hofmann2012, Bello2016, Bello2017a} and experimentally 
\cite{Winkler2006, Folling2007, Strohmaier2010, Preiss2015} 
with cold atoms in optical lattices.

\begin{figure}[t]
    \centering\includegraphics{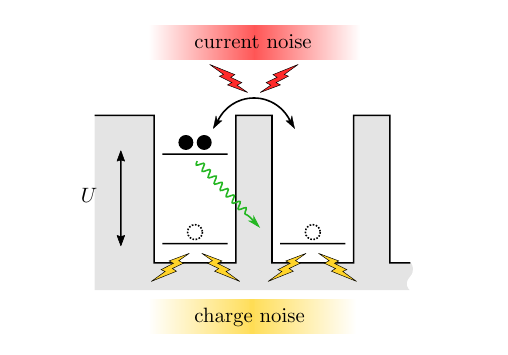}
    \caption{Tight-binding lattice occupied by two electrons.  The
    initial state with a doubly occupied site (doublon) may decay dissipatively
    into a single-occupancy state with lower energy.  The released energy
    is of the order of the onsite interaction $U$ and will be absorbed by
    heat baths representing environmental charge and current noise.
\label{fig:setup}
}
\end{figure}

In the context of solid-state based quantum information and quantum
technologies, arrays of tunnel coupled quantum dots represent a recent platform
for similar experiments with electrons.\cite{Puddy2015a, Zajak2016a}  In
comparison to optical lattices, however, these systems are way more
sensitive to decoherence and dissipation stemming from the interaction with
environmental degrees of freedom such as phonons or charge and current
noise.  Since environments may absorb energy, the separation of two
electrons in a doublon state is no longer energetically forbidden.  In this
paper we cast some light on this issue by studying the life times of
doublons in a one-dimensional lattice in the presence of charge and current
noise, as is sketched in Fig.~\ref{fig:setup}.  For the environment we
employ a Caldeira-Leggett model, \cite{Leggett1987a, Hanggi1990a} where
depending on the type of noise, the bath couples locally to the onsite
energies or to the tunnel matrix elements.

In Sec.~\ref{sec:model}, we specify our model and sketch the derivation of
a Bloch-Redfield master equation for the dissipative dynamics.  Section
\ref{sec:charge} is devoted to the influence of charge noise, while the
results for current noise are worked out in Sec.~\ref{sec:current}.
Boundary effects and experimental consequences are discussed in
Sec.~\ref{sec:discussion}, while the appendix contains details of the
master equation and the averaging of decay rates.

\section{Model and master equation}
\label{sec:model}

The Fermi-Hubbard model considers particles on a lattice with nearest
neighbor tunneling and onsite interaction.  For electrons, its
Hamiltonian reads
\begin{align}
    H_S & = -J\sum\limits_{j=1}^{N-1}\sum\limits_{\sigma=\uparrow, \downarrow}
    \left( c^\dagger_{j+1\sigma}c_{j\sigma} + \mathrm{H.c.}\right)  
+ U\sum\limits_{j=1}^{N} n_{j\uparrow}n_{j\downarrow} \nonumber \\
& \equiv -J T + UD \,,
\label{eq:hubbard}
\end{align}
with the hopping matrix element $J$ and the interaction strength $U$.  The
fermionic operator $c^\dagger_{j\sigma}$ creates an electron with spin
$\sigma$ on site $j$, while $n_{j\sigma}$ is the corresponding number
operator.  For convenience, we define the hopping operator between 
sites $j$ and $j+1$, as $T_j = \sum_\sigma c^\dagger_{j+1\sigma}c_{j\sigma} +
\mathrm{H.c.}$
While the Hamiltonian~\eqref{eq:hubbard} has open boundary conditions,
we will also study the case of periodic boundary conditions (ring
configuration) by adding the corresponding term for the hopping between
the first and the last site.

Henceforth, we focus on the case of two fermions forming a spin singlet.
Then we work in a Hilbert space that contains two types of states,
\textit{single-occupancy states}
\begin{equation}
\label{single}
 \frac{1}{\sqrt{2}} (c^\dagger_{i\uparrow}c^\dagger_{j\downarrow}-
 c^\dagger_{i\downarrow}c^\dagger_{j\uparrow})\ket{0} \ , 
 \quad 1\leq i < j \leq N \ , 
\end{equation}
and the \textit{double-occupancy states}, known as \textit{doublons},
\begin{equation}
\label{double}
 c^\dagger_{j\uparrow}c^\dagger_{j\downarrow}\ket{0} \ , \quad j=1,\cdots,N \,. 
\end{equation}
Both kinds of states are eigenstates of the operator $D$, which in the
Hilbert space considered is equal to the projector onto the doublon
states~\eqref{double}, in the following denoted as $P_D$.

While being different from the states in Eqs.~\eqref{single} and
\eqref{double}, for sufficiently large values of $U$, the eigenstates of 
$H_S$ also discern into two groups, namely $N(N-1)/2$ states with energies 
$|\epsilon_n|\lesssim 4J$ and $N$ states, with energies $|\epsilon_n|\approx U$.
We will refer to the two groups as the \textit{low-energy subspace} 
$\mathcal{H}_0$, and the span of the latter as the \textit{high-energy 
subspace} $\mathcal{H}_1$. In the strongly-interacting regime with $U\gg J$,
treating the tunneling term as a perturbation, it is possible to express the 
projector onto the high-energy subspace $P_1$ as a power series in $J/U$,
see Ref.~\onlinecite{MacDonald1988},
\begin{equation}
    P_1 = P_D - \frac{J}{U}(T^+ + T^-) + 
    \mathcal{O}\left( \frac{J^2}{U^2} \right) \ , \label{eq:projector}
\end{equation}
where ${T^+=P_DT(\mathbb{I}-P_D)}$ and ${T^-=(\mathbb{I}-P_D)TP_D}$
comprise the hopping processes that increase and decrease the double
occupancy respectively. $\mathbb{I}$ is the identity operator.

A key ingredient to our model is the coupling to environmental degrees of
freedom described as $N$ independent baths of harmonic oscillators,
\cite{Leggett1987a,Hanggi1990a}
${H_B=\sum_{j,n}\omega_n a^\dagger_{jn} a_{jn}}$.  They couple to the
Fermi-Hubbard chain via the Hamiltonian ${H_{SB}=\sum_j X_j \xi_j}$,
where the $X_j$ are system operators that will be specified below.
For ease of notation, we introduce the collective bath coordinates
${\xi_j=\sum_n g_n (a^\dagger_{jn}+a_{jn})}$.  Moreover, we assume that all
baths are equal and statistically independent, such that
${\mean{\xi_i(t),\xi_j(t')}= 2S(t-t')\delta_{ij}}$.

Assuming weak coupling and Markovianity, the time evolution of the 
system's density matrix $\rho$, can be suitably described by a master equation 
of the form \cite{Redfield1957a,Breuer2007}
\begin{align}
\label{BlochRedfield}
    \dot{\rho} & = -i[H_S,\rho]-\sum_j [X_j,[Q_j,\rho]]- 
    \sum_j [X_j,\{R_j,\rho\}] \\
    & \equiv -i[H_S,\rho] + \mathcal{L}[\rho] \ .
    \nonumber
\end{align}
with the operators
\begin{align}
  Q_j & = \frac{1}{\pi}\int_0^\infty \diff\tau\int_0^\infty \diff\omega 
  \mathcal{S}(\omega)\tilde{X_j}(-\tau) \cos \omega\tau \ , \label{eq:R}\\
  R_j &= \frac{-i}{\pi}\int_0^\infty \diff\tau\int_0^\infty \diff\omega 
  \mathcal{J}(\omega)\tilde{X_j}(-\tau) \sin \omega\tau \ . \label{eq:Q}
\end{align}
The tilde denotes the interaction picture with respect to the system
Hamiltonian, ${\tilde{X}_j(-\tau)=e^{-i H_S\tau}X_je^{i H_S\tau}}$, while
${\mathcal{J}(\omega)=\pi\sum_n |g_n|^2 \delta(\omega- \omega_n)}$ is the
spectral density of the baths and
${\mathcal{S}(\omega)=\mathcal{J}(\omega)\coth(\beta\omega/2)}$ is the
Fourier transformed of the symmetrically-ordered equilibrium
autocorrelation function $\mean{\{\xi_j(\tau),\xi_j(0)\}}/2$.
$\mathcal{J}(\omega)$ and $\mathcal{S}(\omega)$ are independent of $j$
since all baths are identical. We will assume an ohmic spectral density
${\mathcal{J}(\omega)=\pi\alpha\omega/2}$, where the dimensionless parameter
$\alpha$ characterizes the dissipation strength.

\section{Charge noise}
\label{sec:charge}

Fluctuations of the background charges in the substrate essentially act
upon the charge distribution of the chain. Therefore, we model it by
coupling the occupation of each site to a heat bath, such that
\begin{equation}
H_{SB}^Q = \sum_{j,\sigma} n_{j,\sigma} \xi_j \,,
\end{equation}
which means $X_j = n_j$. This fully specifies the master equation~\eqref{BlochRedfield}.

\begin{figure}
    \centering\includegraphics{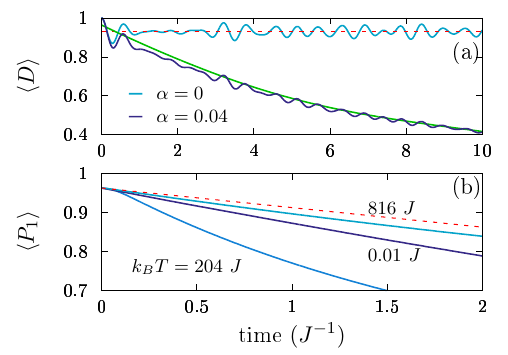}
    \caption{Time evolution of the double occupancy in a system with charge
        noise. The initial state consists of a doublon localized in a 
        particular site of a chain with periodic boundary conditions.  
        Parameters: $N=5$, $U=10J$ and $\alpha=0.04$. (a) Comparison between 
        free dynamics ($\alpha=0$) and dissipative dynamics ($\alpha\neq 0$). 
        Temperature is set to $k_B T=0.01J$. The green line corresponds to the 
        occupancy of the high-energy subspace for the case with $\alpha\neq 0$ 
        and illustrates the bound given in \eqref{eq:bound}. (b) Decay of the 
        high-energy subspace occupancy for different temperatures ranging from  
        $0.01J$ to $1000J$. The slope of the curves at time $t=0$ is the same 
        in all cases and coincides with the value given by \eqref{eq:avergamma} 
        (red dashed line).} 
        \label{fig:dynamics} 
\end{figure}

To get a qualitative impression of the decay dynamics of a doublon, let us
start by discussing the time evolution of a doublon state in the strongly
interacting regime shown in Fig.~\ref{fig:dynamics}.  For $\alpha=0$,
i.e., in the absence of dissipation, the two electrons will essentially
remain together throughout time evolution. This is due to energy
conservation and the fact that kinetic energy in a lattice is bounded, it
can be at most $2|J|$ per particle. Thus, particles forming a doublon
cannot split, as they would not have enough kinetic energy on their own to
compensate for the large $U$.  However, since the doublon states are not
eigenstates of the system Hamiltonian, we observe some slight oscillations
of the double occupancy $\langle D\rangle$.  Still the time average of
this quantity stays close to unity, see Fig.~\ref{fig:dynamics}(a). 

On the contrary, if the system is coupled to a bath, doublons will be able to 
split releasing energy into the environment.  Then the density operator
eventually becomes the thermal state $\rho_\infty\propto e^{-\beta H_S}$.
Depending on the temperature and the interaction strength, the
corresponding asymptotic doublon occupancy $\mean{D}_\infty$ may still
assume an appreciable value.

\subsection{Numerical analysis \label{sec:numerics}}

To gain quantitative insight, we decompose our master equation
\eqref{BlochRedfield} into the system eigenbasis and obtain a form
convenient for numerical treatment (for details, see
Appendix~\ref{app:masterequation}). A typical time evolution of the occupancy
$\langle D\rangle$ is shown in Fig.~\ref{fig:dynamics}(a).  It exhibits an
almost mono-exponential decay, such that the doublon life time $T_1$ can be
defined as the $1/e$ time of the difference between initial and final
value of $\langle D\rangle$,
\begin{equation}
    \frac{\mean{D}_{T_1}-\mean{D}_\infty}{1-\mean{D}_\infty}=\frac{1}{e} \,.
    \label{eq:numericgamma}
\end{equation}

\begin{figure}[tb] 
    \centering\includegraphics{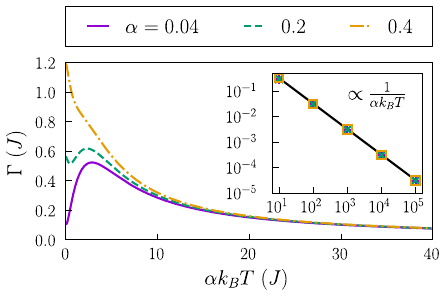} 
    \caption{Temperature dependence of the numerically obtained decay rate 
        for a chain with $N=5$ and periodic boundary conditions in the presence
        of charge noise. The interaction energy is set to $U=10J$. For values 
        of the coupling strength $\alpha\lesssim 0.04$ we obtain approximately 
        the same curve (continuous line). Inset: Values for $\Gamma$ on a 
        logarithmic scale demonstrating the proportionality 
        $\propto 1/\alpha T$.} 
\label{fig:numerics_occup}
\end{figure}

The corresponding decay rate $\Gamma = 1/T_1$ is shown in
Fig.~\ref{fig:numerics_occup} as a function of the temperature for different
values of the dissipation strength $\alpha$. For small $\alpha$ and
intermediate temperatures, $\Gamma$ increases with the temperature,
reaching a maximum after which the tendency inverts.  For sufficiently
large temperatures, $\Gamma \propto (\alpha k_B T)^{-1}$.

\subsection{Analytical estimate for the decay rate}

An analytical estimate for the decay rates can often be gained from the
behavior at the initial time $t=0$, i.e.\ from $\dot\rho(0) =
-i[H_S,\rho_0]+ \mathcal{L}\rho_0$ with $\rho_0=\rho(0)$ being the pure
initial state.  In the present case, however, the calculation is hindered
by the fast initial oscillations witnessed in Fig.~\ref{fig:dynamics}(a).
These oscillations stem from the mixing of the doublon states with the
single-occupancy states.  To circumvent this problem, we focus for the
present purpose on the occupancy of the high-energy subspace, $\langle
P_1\rangle$ shown in Fig.~\ref{fig:dynamics}(b).  It turns out that this
quantity evolves more smoothly while it decays also on the time scale
$T_1$.  The reason for its lack of fast oscillations is that the projector
$P_1$ commutes with the system Hamiltonian, so that it expectation value is
determined solely by dissipation.  Notice that the initial decay is
temperature independent, while at a later stage, the decay is strongest for
intermediate temperatures.

A formal way of understanding the similarity of the long time dynamics of
$\langle D\rangle$ and $\langle P_1\rangle$ is provided by the estimate
\begin{align}
    |\tr{P_1\rho}-\tr{D\rho}|
    \leq {}& \sqrt{2}\|\rho\|\sqrt{N-\tr{P_1P_D}} \\
    \simeq {}& 2\sqrt{2N}J/U \ ,
    \label{eq:bound}
\end{align}
where the first lines follows from the Cauchy-Schwarz inequality for the
inner product of operators, $(A,B)=\tr{A^\dagger B}$,
while the second line stems from the
perturbative expansion of $P_1$ given by Eq.~\eqref{eq:projector}.  The
result implies that when neglecting corrections of the order of $J/U$, we may
determine $T_1$ and $\Gamma$ from either quantity.  Nevertheless it is
instructive to analytically evaluate $\Gamma$ for the decay of both
$\langle D\rangle$ and $\langle P_1\rangle$.

Following our hypothesis of a mono-exponential decay, we expect
\begin{equation}
    \label{eq:P1decay}
    \mean{P_1}\simeq \Delta e^{-\Gamma t} + \mean{P_1}_\infty \ ,
\end{equation} 
therefore,
\begin{equation}
    \Gamma\simeq - \frac{1}{\Delta}
    \left. \frac{\diff \mean{P_1}}{\diff t}\right|_{t=0}=
    -\frac{\tr{P_1\mathcal{L}[\rho_0]}}{\mean{P_1}_0-\mean{P_1}_\infty} \ .
    \label{eq:gamma}
\end{equation}
This expression still depends slightly on the specific choice of the
initial doublon state, in particular for open boundary conditions (see
Sec.~\ref{sec:boundary}, below).  To obtain a more global picture, we
consider an average over all doublon states, which can be performed
analytically.\cite{Storcz2005a}  From Eq.~\eqref{eq:gamma}, we find the
average decay rate
\begin{multline}
    \overline{\Gamma}=\frac{1}{N\Delta}\sum_j \tr{P_D[Q_j,[X_j,P_1]]} \\
    -\tr{P_D\{R_j,[X_j,P_1]\}} \ .
    \label{eq:avergamma}
\end{multline}
For details of the averaging procedure, see Appendix~\ref{app:average}.

For a further simplification, we have to evaluate the expressions
\eqref{eq:R} and \eqref{eq:Q} which is possible by approximating the
interaction picture coupling operator as $\tilde{X}_j(-\tau)\simeq X_j
-i\tau[H_S,X_j]$.  This is justified as long as the decay of the
environmental excitations is much faster than the typical system evolution,
i.e., in the high-temperature regime (HT).  Inserting our approximation for
$\tilde{X}_j$ and neglecting the imaginary part of the integrals, we arrive at
\begin{align} 
    Q_j & \simeq \frac{1}{2}\lim_{\omega\rightarrow 0^+} 
    \mathcal{S}(\omega) X_j = \frac{\pi}{2}\alpha k_B T X_j\ , \\ R_j & \simeq -
    \frac{1}{2}\lim_{\omega\rightarrow 0^+} \mathcal{J}'(\omega) [H_S,X_j] = 
    \frac{\pi}{4}\alpha [H_S,X_j] \ .  
\end{align} 
With these expressions, Eq.~\eqref{eq:avergamma} results in a temperature
independent decay rate.  Notice that any temperature dependence stems from
the $Q_j$ in the first term of Eq.~\eqref{eq:avergamma} which vanishes in
the present case.  While this observation agrees with the numerical findings in
Fig.~\ref{fig:dynamics} for very short times, it does not reflect the
temperature dependent decay of $\langle P_1\rangle$ at the more relevant
intermediate stage.

This particular behavior hints at the mechanism of the bath-induced doublon
decay.  Let us notice that the coupling to charge noise, $X_j=n_j$,
commutes with $D$.  Therefore, the initial state is robust against the
influence of the bath.  Only after mixing with the single-occupancy
states due to the coherent dynamics, the system is no longer in an
eigenstate of the $n_j$, such that decoherence and dissipation become
active.  Thus, it is the combined action of the system's unitary evolution
and the effect of the environment which leads to the doublon decay.

An improved estimate of the decay rate, can be calculated by averaging the
transition rate of states from the high-energy subspace to the low-energy
subspace.  Let us first focus on regime $k_BT\gtrsim U$ in which we can
evaluate the operators $Q_j$ in the high-temperature limit.  Then the
average rate can be computed using expression \eqref{eq:avergamma} and
replacing $P_D$ by $P_1$, see Appendix \ref{app:masterequation}. With the
perturbative expansion of $P_1$ in Eq.~\eqref{eq:projector} we obtain to
leading order in $J/U$ the averaged rate
\begin{equation} 
    \overline{\Gamma}_\mathrm{HT} \simeq 
    \frac{4\pi\alpha J^2}{U^2 \Delta}\left(2k_B T + U \right) \,,
    \label{eq:GOHT}
\end{equation}
valid for periodic boundary conditions.  For open boundary conditions, the
rate acquires an additional factor $(N-1)/N$.  Notice that we have
neglected back transitions via thermal excitations from singly occupied
states to doublon states.  We will see that this leads to some smaller
deviations when the temperature becomes extremely large.  Nevertheless, we
refer to this case as the high-temperature limit.

In the opposite limit, for temperatures ${k_B T < U}$, the decay rate
saturates at a constant value. To evaluate $\overline{\Gamma}$ in this limit, it
would be necessary to find an expression for $\tilde{X}_j(-\tau)$ dealing
properly with the $\tau$-dependence for evaluating the noise kernel, a
formidable task that may lead to rather involved expressions.
Nevertheless, one can make some progress by considering the transition of
one initial doublon to one particular single-occupancy state.  This
corresponds to approximating our two-particle lattice model by the
dissipative two-level system for which the decay rates in the Ohmic case
can be taken from the literature,\cite{Weiss1989a, Makhlin2001b} see
Appendix \ref{app:TLS}.  Relating $J$ to the tunnel matrix element of the
two-level system and $U$ to the detuning, we obtain from
Eq.~\eqref{app:Gammaii} the temperature-independent expression
\begin{equation}
    \overline{\Gamma}_\text{LT}  \simeq \frac{8\pi\alpha J^2}{U \Delta} \ ,
\label{eq:GOLT}
\end{equation}
which formally corresponds to Eq.~\eqref{eq:GOHT} with the temperature set
to $k_B T=U/2$.

\begin{figure}[tb]
    \centering\includegraphics{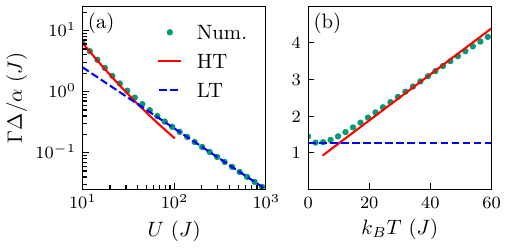}
    \caption{Comparison between the numerically computed decay rate and the 
        analytic formulas \eqref{eq:GOHT} and \eqref{eq:GOLT} for a chain
        with $N=5$ sites and periodic boundary conditions in the case of 
        charge noise.  The dissipation strength is $\alpha=0.02$.
        (a) Dependence on the interaction strength for a fixed temperature $k_B 
        T=20J$. (b) Dependence on the temperature for a fixed interaction 
        strength $U=20J$.}
    \label{fig:analyticsOccup}
\end{figure}

Figure~\ref{fig:analyticsOccup} provides a comparison of these analytical
findings with numerical results.  The data in panel (a) reveal that the
transition between the low-temperature regime and the high-temperature
regime is rather sharp and occurs at $U\approx k_BT$.  Panel (b) shows
$\Gamma$ as a function of the temperature.  For low temperatures, the
numerical values saturate at $\overline{\Gamma}_\text{LT}$ obtained from the
approximate mapping to a two-level system.  For high temperatures, the
analytical prediction $\overline{\Gamma}_\text{HT}$ seems slightly too 
large.  The discrepancy stems from neglecting thermal excitations, as 
mentioned above.

\section{Current noise}
\label{sec:current}

\begin{figure}
    \centering\includegraphics{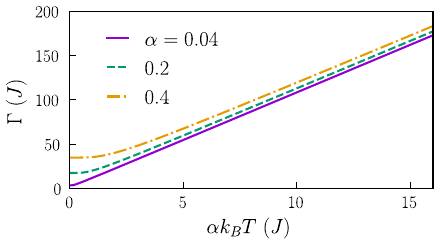}
    \caption{Average decay rate of the doublon states under the influence
    of current noise for various dissipation strengths as a function of
    the temperature.  The chain consists of $N=5$ sites with periodic boundary
    conditions, while the interaction is $U=10J$.}
    \label{fig:numerics_hopp}
\end{figure}

Fluctuating background currents mainly couple to the tunnel matrix elements
of the system.  Then the system-bath interaction is given by setting 
$X_j=T_j$ and reads
\begin{equation}
H_{SB}^I = \sum_{j,\sigma} (c^\dagger_{j+1\sigma}c_{j\sigma} 
+ c^\dagger_{j\sigma} c_{j+1\sigma} ) \xi_j \,.
\end{equation}
Depending on the boundary conditions, the sum may include the term with
$j=N$.  The main qualitative difference of this choice is that in contrast
to charge noise, $H_{SB}^I$ does not commute with the projector to the
doublon subspace and, thus, generally ${\tr{D\mathcal{L}[\rho]}\neq 0}$.
This enables a direct dissipative decay without the detour via an admixture
of single-occupancy states to the doublon states.  As a consequence,
for the same value of the dimensionless dissipation parameter $\alpha$, the
decay may be much faster.  Also the temperature dependence of the decay
changes significantly, as can bee seen in Fig.~\ref{fig:numerics_hopp}.
While $\overline{\Gamma}$ is still proportional to $\alpha$, it now grows
monotonically with the temperature.

\begin{figure}
    \centering\includegraphics{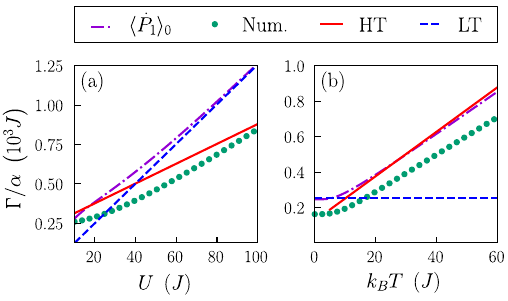}
    \caption{Numerically obtained decay rate in comparison with the
        approximations~\eqref{eq:decay_hopp}, \eqref{eq:GTHT} and 
        \eqref{eq:GTLT} for a chain with $N=5$ sites and periodic boundary 
        conditions in the case of current noise with strength $\alpha=0.02$.  
        The results are plotted as a function of (a) the interaction and the 
        temperature $k_B T=20J$ and (b) for a fixed interaction $U=20J$ as a 
        function of the temperature.}
    \label{fig:analyricsHopp}
\end{figure}

As in the last section, we proceed by calculating analytical estimates for
the decay rates.  However, since the time evolution is no longer
mono-exponential (not shown), we no longer start from the
ansatz~\eqref{eq:avergamma}, but estimate the rate from the slope of the
occupancy $\langle P_1\rangle$ at initial time,
\begin{equation}
    \Gamma \simeq - \left. \frac{\diff \mean{P_1}}{\diff t} \right|_{t=0} =
        - \tr{P_1\mathcal{L}[\rho_0]} \ .
    \label{eq:decay_hopp}
\end{equation}
We again perform the average over all doublon states for $\rho_0$
in the limits of high and low temperatures.  For periodic boundary
conditions, we obtain to lowest order in $J/U$ the high and low temperature
rates 
\begin{align}
    \overline{\Gamma}_\mathrm{HT} & = 2\pi\alpha \left(2k_B T+U \right) \,,
    \label{eq:GTHT}  \\
    \overline{\Gamma}_\text{LT} & = 4\pi\alpha U \,,
    \label{eq:GTLT}
\end{align}
respectively,
while open boundary conditions lead to the same expressions but with a
correction factor $(N-1)/N$.
In Fig.~\ref{fig:analyricsHopp}, we compare these results with the
numerically evaluated ones as a function of the interaction
[Fig.~\ref{fig:analyricsHopp}(a)] and
the temperature [Fig.~\ref{fig:analyricsHopp}(b)].  Both show that the
analytical approach correctly predicts the (almost) linear behavior at
large values of $U$ and $k_BT$, as well as the saturation for small values.
However, the approximation slightly overestimates the influence of the
bath.

While the rates reflect the decay at short times, it is worthwhile to
comment on the long time behavior under the influence of current
noise.  For open chains as well as for closed chains with an even
number of sites, it is not ergodic as the long-time solution is not unique.
The reason for this is the existence of a doublon state
${\ket{\Phi}=\frac{1}{\sqrt{N}}\sum_{j=1}^N (-1)^j
c^\dagger_{j\uparrow}c^\dagger_{j\downarrow}\ket{0}}$ which is an
eigenstate of $H_S$ without any admixture of single-occupancy states.
Since $T_j\ket{\Phi}=0$ for all sites $j$, current noise may affect
the phase of ${\ket{\Phi}}$, but cannot induce its dissipative decay.
For a closed chain with an odd number of sites, by contrast, the alternating
phase of the coefficients of $\ket{\Phi}$ is incompatible with periodic
boundary conditions, unless a flux threads the ring.  As a consequence,
the chain eventually resides in the thermal state $\propto\exp(-\beta
H_S)$.  The difference is manifest in the final value of the doublon
occupancy at low temperatures.  For closed chains with an odd number of
sites, it will fully decay, while in the other cases, the population of
$\ket{\Phi}$ will survive.

\section{Discussion}
\label{sec:discussion}
\subsection{Dimension and boundary effects}
\label{sec:boundary}

\begin{figure}[tb]
    \centering\includegraphics{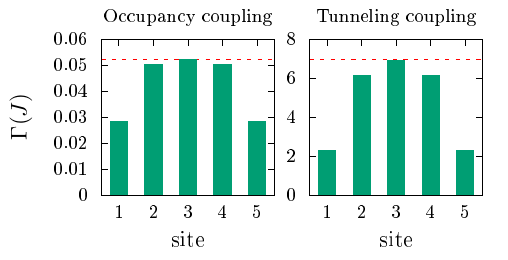}
    \caption{Decay rates of the double occupancy for a chain with $N=5$ sites 
        with open boundary as a function of the initial location of the doublon.
	The values for $\Gamma$ are taken as the inverse of the $T_1$ time
	obtained from a numerical propagation of the master equation.
        The red dashed line marks the value for closed boundary conditions.
	The other parameters are $U=20J$, $\alpha=0.01$, $k_B T=5J$.}
    \label{fig:site_dependence}
\end{figure}

So far, we have considered decay rates as the averages of all possible
initial doublon or high-energy states.  While this is sufficient for a
generic estimate of the life times, it ignores the fact that the behavior
of individual states may differ significantly, in particular when the
initial state is located at a boundary, which reduces the number of
accessible decay channels.  In Fig.~\ref{fig:site_dependence} we present
the decay rates for doublons as a function of the initial site.  It reveals
that in comparison to states at the center, an initial localization at the
first or last site, may double the life time for charge noise and enhance by
it by a factor three for current noise.  The dashed lines in these plots
marks the value for periodic boundary conditions, for which the value is
practically the same as for a states in the center.

This knowledge about the role of boundaries and nearest neighbors provides
some hint on the doublon life time in higher-dimensional lattices.  Let us
notice that Let the decay rates \eqref{eq:gamma} and \eqref{eq:decay_hopp} 
contain one term for each single-occupancy state that is directly tunnel 
coupled to the initial site.  Assuming that all terms are of the same order, 
we expect that $\overline{\Gamma}$ is by and large proportional to the 
coordination number of the lattice sites.  Therefore the life time should 
decrease only moderately with the dimension, roughly as 
$T_1 = \Gamma^{-1} \sim 2^{-D}$.  From the data in 
Fig.~\ref{fig:site_dependence}(b), we can appreciate that for current
noise, the difference between center and border is even larger.  Thus,
increasing dimensionality should have a slightly larger impact on the
doublon life times.

\subsection{Experimental implications}
\label{sec:experiment}

A current experimental trend is the fabrication of larger arrays of quantum
dots,\cite{Puddy2015a, Zajak2016a} which triggered our question on the
feasibility of doublon experiments in solid-state systems.  While the size
of these arrays would be sufficient for this purpose, their dissipative
parameters are not yet fully known.  For an estimate we therefore consider
the values for GaAs/InGaAs quantum dots which have been determined recently
via Landau-Zener interference.\cite{Forster2014, onalpha}
Notice, that for the strength of the current noise, only an upper
bound has been reported.  We nevertheless use this value, but keep in mind
that it leads to a conservative estimate.  In contrast to the former
sections, we now compute the decay for the simultaneous action of charge
noise and current noise.

\begin{figure}
    \centering\includegraphics{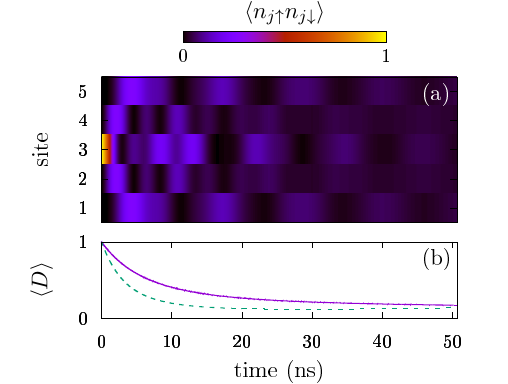}
    \caption{(a) Spatially resolved doublon dynamics in a chain with $N=5$ 
    sites and open boundary conditions for the dissipative parameters
    determined in Ref.~\onlinecite{Forster2014}, i.e., for the dissipation
    strengths\cite{onalpha} $\alpha_Q=3\times 10^{-4}$ and $\alpha_I=5\times 10^{-6}$, 
    the tunnel coupling $J=13\,\mu\mathrm{eV}$, interaction $U=1.3$\,meV, and 
    temperature $T=10$\,mK.  (b) Corresponding decay of the double occupancy 
    (solid line) and state purity (dashed).}
    \label{fig:realistic_dynamics}
\end{figure}

\begin{figure}
    \centering\includegraphics{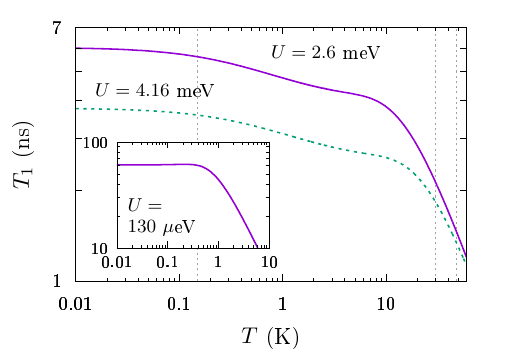}
    \caption{Doublon life time as a function of the temperature for
	different interaction strengths.  The other parameters are as in
	Fig.~\ref{fig:realistic_dynamics}. Vertical dashed lines mark the 
    temperature corresponding to the tunneling energy and the Hubbard 
    interaction energy.  
    Inset: $T_1$ time for the optimized value of the interaction,
	$U =10J =130\,\mu$eV and a current noise with
	$\alpha_I = 2\times10^{-6}$.  The latter is smaller than the value in
	Fig.~\ref{fig:realistic_dynamics}, but still realistic.
	}
    \label{fig:realistic_experiment}
\end{figure}

Figure~\ref{fig:realistic_dynamics}(a) shows the dissipative time evolution
for a doublon initially localized at the center of a chain with 5 sites.
The dynamics exhibits a few coherent oscillations in which the doublon
evolves into a superposition of the kind $|2,0,0\rangle + |0,0,2\rangle$,
which represents an example of a NOON state.\cite{Lee2002a}
Each component propagates to one end of the chain, where is it reflected
such that subsequently the initial states revives.  In
Fig.~\ref{fig:realistic_dynamics}(b), we depict the evolution of the
corresponding doublon occupancy and the purity.  Both quantities decay
rather smoothly.  This agrees to the finding found in
Sec.~\ref{sec:current} for pure current noise which obviously dominates.
It is also consistent with the values for the respective analytical decay
rates in the low-temperature limit.  Figure~\ref{fig:realistic_experiment}
shows the $T_1$ times for two different interaction strengths.  It reveals
that for low temperatures $T\lesssim J/k_BT$, the life time is essentially
constant, while for larger temperatures, it decreases moderately until
$k_BT$ comes close to the interaction $U$.  For higher temperatures,
$\Gamma$ starts to grow linearly.  On a quantitative level, we expect life
times of the order $T_1\sim 5$\,ns already for a moderately low
temperatures $T\lesssim 100$\,mK.  Since we employed the value of the upper
bound for the current noise, the life time might be even larger.

Considering the analytical estimates for the decay rates at low
temperatures, Eqs.~\eqref{eq:GOLT} and \eqref{eq:GTLT}, separately,
lets us conclude that for smaller values of $U$, current noise becomes
less important, while the impact of charge noise grows.  Therefore, a
strategy for reaching larger $T_1$ times is to
design quantum dots arrays with smaller onsite interaction, such
that the ratio $U/J$ becomes more favorable.  The largest $T_1$ is expected
in the case in which both low-temperature decay rates are equal,
$\overline{\Gamma}_\text{LT,charge} = \overline{\Gamma}_\text{LT,current}$, 
which for the present experimental parameters is found at $U\sim 10J$ (while 
our data is for $U\sim 100 J$).  This implies that in an optimized device, 
the doublon life times could be larger by one order of magnitude to reach 
values of $T_1\sim 50$\,ns, which is corroborated by the data in the inset 
of Fig.~\ref{fig:realistic_experiment}.

\section{Conclusions}

We have investigated the life times of double-occupancy states or doublons
in a one-dimensional Hubbard model under the influence of dissipating
environments.  While in optical lattices, the resulting dissipative decay
may be of minor influence, for quantum dot arrays, it will be a limiting
factor.

We have considered two different couplings between the system and its
environment, which physically correspond to the impact of charge noise and
current noise, respectively.  Within a Bloch-Redfield formalism, this model
can be treated with a master equation, which allows one to
numerically determine the life times from the time evolution of the reduced
density operator.  Moreover, it provides analytical estimates for the
initial decay rates.  It turned out that the striking difference between
the two couplings is that the impact of charge noise decreases with the
interaction, while current noise becomes increasingly relevant.

For present quantum dots, the doublon life time is expected to be of the
order 5\,ns, which would limit the coherent dynamics to only a few
periods.  However, our analytical estimates suggest that for quantum dot
arrays with smaller onsite interaction, an extension by one order of
magnitude should be feasible.  Thus, the recent trend towards arrays with
ever more coherently coupled quantum dots will allow the experimental
realization of effects that so far have been measured only in optical
lattices.

\appendix
\section{Master equation in the system eigenbaasis \label{app:masterequation}}

To bring the master equation \eqref{BlochRedfield} into a form that is
suitable for a numerical implementation, we have to evaluate the
$\tau$-integrals in Eqs.~\eqref{eq:R} and \eqref{eq:Q}.  This is possible
after a decomposition into the system eigenbasis $\{\ket{\phi_\alpha}\}$
with $H_S\ket{\phi_\alpha} =\epsilon_\alpha\ket{\phi_\alpha}$.  Then the
transformation to the interaction picture provides phase factors yielding a
Dirac delta function and a principal value integral. Neglecting the latter,
as it usually consist in a renormalization of the free system parameters,
and using the notation ${\rho_{\alpha\beta}\equiv\bra{\phi_\alpha} \rho
\ket{\phi_\beta}}$ and ${X^{(j)}_{\alpha\beta}\equiv\bra{\phi_\alpha} X_j
\ket{\phi_\beta}}$, the master equation becomes
\begin{equation}
\dot{\rho}_{\alpha\beta}=-i(\epsilon_\alpha-\epsilon_\beta)\rho_{\alpha\beta} 
+ \sum_{\alpha'\beta'}\mathcal{L}_{\alpha\beta,\alpha'\beta'} 
\rho_{\alpha'\beta'} \,. \label{eq:mastereq}
\end{equation}
The generalized golden-rule rates
\begin{equation}
\begin{split}
\mathcal{L}_{\alpha\beta,\alpha'\beta'} ={}& \sum_j \Big[(\Gamma_{\beta'\beta}
    +\Gamma_{\alpha'\alpha})X^{(j)}_{\alpha\alpha'}X^{(j)}_{\beta'\beta}
    \\ &
    - \delta_{\beta\beta'}\sum_{\beta''}
    \Gamma_{\alpha'\beta''}X^{(j)}_{\alpha\beta''}X^{(j)}_{\beta''\alpha'}
    \\ &
    - \delta_{\alpha\alpha'}\sum_{\alpha''}
    \Gamma_{\beta'\alpha''}X^{(j)}_{\beta'\alpha''}X^{(j)}_{\alpha''\beta} 
    \Big] \,,
\end{split}
\end{equation}
are determined by the transition matrix elements of the system operator
that couples to the bath and the factors
${\Gamma_{\alpha \beta} \equiv \Gamma(\epsilon_\alpha-\epsilon_\beta)}$
with
\begin{equation}
\Gamma(\omega)=\begin{cases} J(\omega)(1+n_B(\omega)) & \omega>0 \\
                             J(-\omega)n_B(-\omega) & \omega<0
\end{cases} \,,
\end{equation}
and the thermal bosonic occupation number $n_B(\omega)=(e^{\beta\omega}-1)^{-1}$. 

The Bloch-Redfield equation allows the direct computation of decay rates
averaged over all possible initial states, which in our case are doublon
states or high-energy states.  To this end, we distinguish, those from a
set $I_1$ labeling the high-energy states and $I_0$ for the low-energy
states.  With the formulas for the averages derived in the
Appendix~\ref{app:average} and the projector to the high-energy subspace
$P_1$, we arrive at
\begin{multline}
    \overline{\Gamma}=\frac{1}{N\Delta}\sum_j \tr{P_1[Q_j,[X_j,P_1]]} \\
    -\tr{P_1\{R_j,[X_j,P_1]\}} \,.
    \label{eq:decay_occup}
\end{multline}
Notice that the factor $\Delta$ accounts for the finite final value of the
decay in Eq.~\eqref{eq:P1decay}.  Therefore,
\begin{equation}
\label{rateA}
    \overline{\Gamma}=-\frac{1}{\Delta}\frac{\diff\overline{\mean{P_1}}}{\diff
    t}\Big|_{t=0}
    =-\frac{1}{\Delta} \overline{\tr{P_1\mathcal{L}[\rho]}} \ , 
\end{equation}
where the bar denotes the average over all pure states belonging to the
high-energy subspace, instead of the doublon subspace, see Appendix
\ref{app:average}. An alternative form for this quantity is
\begin{align}
\label{rateAB}
 \overline{\Gamma}\Delta
= -\frac{1}{N}\sum_{\alpha,\beta\in I_1}\mathcal{L}_{\alpha\alpha,\beta\beta} 
= \frac{1}{N}\sum_{\alpha\in I_0} \sum_{\beta\in I_1}\mathcal{L}_{\alpha\alpha,\beta\beta}\ ,
\end{align}
where the last equality follows from the trace preserving property
of the master equation, 
$\sum_\alpha \mathcal{L}_{\alpha\alpha,\beta\beta} = 0$.

\section{Average over pure initial states \label{app:average}}

As an ensemble of pure states, we consider any linear combination
$\ket{\psi}=\sum_{n=1}^N c_n \ket{n}$ of orthonormal basis states
$|n\rangle$, ${n=1,\ldots,N}$.  As a minimal requirement, we postulate that
$\ket{\psi}$ is normalized and invariant under unitary transformations.
Then its probability density reads
\begin{equation}
P(c_1,\cdots,c_N)=\frac{(N-1)!}{\pi^N}\delta\left(1-r^2\right) \,,
\end{equation}
where $r^2=\sum_{n=1}^N |c_n|^2$. Then averages of the 
kind $\overline{c_n c^*_m}$ or $\overline{c_n c^*_m c_{n'} c^*_{m'}}$ can
be expressed as integrals of polynomials over the $(2N-1)$-dimensional 
unit sphere.  Following Ref.~\onlinecite{Folland2001}, we find
\begin{align}
 \overline{c_n c^*_m} & =\frac{1}{N}\delta_{nm} \ ,\\
 \overline{c_n c^*_m c_{n'} c^*_{m'}} & = \frac{1}{N(N+1)}(\delta_{nm}\delta_{n'm'}+\delta_{nm'}\delta_{n'm}) \ ,
\end{align}
which implies
\begin{align}
  \overline{\tr{\rho A}} & =\frac{1}{N} \tr{A} \ , \\
  \overline{\tr{\rho A \rho B}} & =\frac{\tr{A}\tr{B}+\tr{AB}}{N(N+1)} \ .
\end{align}

To compute average rates for the transitions between two groups of states,
cf.\ Eq.~\eqref{rateAB}, the initial linear combination $\ket{\psi}$ 
is restricted to the doublon subspace which has dimension $N_D$.  Therefore we
have to replace the prefactor $N$ by $N_D$ and the operators $A$ and $B$ by
their projections to the subspace, $P_DAP_D$ and $P_DBP_D$. 

\section{Two-level system \label{app:TLS}}

For completeness, we summarize the Bloch-Redfield result for the decay
rates of the two-level system coupled to an Ohmic bath.\cite{Weiss1989a,
Makhlin2001b}  For the notation used in the main text, it is defined by the
Hamiltonian
\begin{equation}
  H =\frac{\Delta}{2}\sigma_x+\frac{\epsilon}{2}\sigma_z
     + \frac{1}{2} X \xi \,,
\label{eq:tls}
\end{equation}
with the tunnel matrix element $\Delta$ and the detuning $\epsilon$.  The bath
coupling is specified by (i) $X=\sigma_z$ for charge noise and (ii)
$X=\sigma_x$ for current noise, respectively.
To establish a relation to our Hubbard chain, we identify the detuning by
the interaction, $\epsilon\simeq U$, and $\Delta=\sqrt{8}J$.
Note that replacing charge noise by current noise corresponds to
interchanging $\epsilon$ and $\Delta$.  Therefore, we can restrict the
derivation of the decay rate to case (i).

It is straightforward to transform the Hamiltonian into the eigenbasis of
the two-level system, where it reads
\begin{equation}
 H' = \frac{E}{2}\sigma_z+X\xi \ , 
\end{equation}
with $E=\sqrt{\epsilon^2 + \Delta^2}$, while the system-bath coupling becomes
\begin{equation}
	X'= \frac{\epsilon}{2E}\sigma_x + \frac{\Delta}{2E}\sigma_z \,.
\end{equation}
In the interaction picture, it is
\begin{multline}
 \tilde{X}(-\tau) 
 =\frac{1}{2E}\left(\epsilon\sigma_x\cos E\tau + \epsilon\sigma_y\sin E \tau + \Delta\sigma_z\right) \ .
\end{multline}
Again ignoring the imaginary part of the integral in \eqref{eq:Q}, the
noise kernel can be written as
\begin{equation}
 Q = \frac{\epsilon}{2E}\frac{\mathcal{S}(E)}{2}\sigma_x + \frac{\Delta}{2E}\frac{\mathcal{S}(0)}{2}\sigma_z \ .
\end{equation}
The projector to the initial state is $P_1=(\sigma_0 + \sigma_z)/2$, so
that the decay rate can be found as
\begin{equation}
    \Gamma_\text{i} = \tr{P_1[Q,[X,P_1]]}
    = \left(\frac{\epsilon}{2E}\right)^2\mathcal{S}(E) \,,
    \label{eq:gammatls}
\end{equation}
where for an Ohmic spectral density
\begin{equation}
 \mathcal{S}(E)=2\pi \alpha E \coth(E/2k_B T) \ .
\end{equation}
Accordingly, we find for case (ii) the rate
\begin{equation}
\label{app:Gammaii}
    \Gamma_\text{ii}
    = \left(\frac{\Delta}{2E}\right)^2\mathcal{S}(E) \,,
\end{equation}
which provides the analytical high-temperature result \eqref{eq:GOLT} for
charge noise.

\bibliography{Doublon_decay,literature,footnote}

\end{document}